\documentclass[twocolumn,showpacs,preprintnumbers,amsmath,amssymb,superscriptaddress,prb]{revtex4}

\usepackage{graphicx}
\usepackage{dcolumn}
\usepackage{bm}

\begin{document}

\preprint{}

\title{Spin splitting and Kondo effect in quantum dots coupled to 
noncollinear ferromagnetic leads}

\author{Daisuke Matsubayashi}
\affiliation{Department of Physics, University of Tokyo, 
7-3-1 Hongo, Bunkyo-ku, Tokyo 113-0033, Japan}
\author{Mikio Eto}
\affiliation{\mbox{Faculty of Science and Technology, Keio University,
3-14-1 Hiyoshi, Kohoku-ku, Yokohama 223-8522, Japan}}

\date{July 21, 2006}

\begin{abstract}
We study the Kondo effect in a quantum dot coupled to
two noncollinear ferromagnetic leads.
First, we study the spin splitting
$\delta\epsilon=\epsilon_{\downarrow}-\epsilon_{\uparrow}$
of an energy level in the
quantum dot by tunnel couplings to the ferromagnetic leads,
using the poor man's scaling method.
The spin splitting takes place in an intermediate direction between 
magnetic moments in the two leads.
$\delta\epsilon \propto
p\sqrt{\cos^2(\theta/2)+v^2\sin^2(\theta/2)}$, where $p$ is the
spin polarization in the leads, $\theta$ is the angle between the
magnetic moments, and $v$ is an asymmetric factor of tunnel
barriers ($-1<v<1$). Hence the
spin splitting is always maximal in the
parallel alignment of two ferromagnets ($\theta=0$) and minimal in
the antiparallel alignment ($\theta=\pi$).
Second, we calculate the Kondo temperature $T_{\mathrm{K}}$.
The scaling calculation yields an analytical expression of
$T_{\mathrm{K}}$ as a function of $\theta$ and $p$,
$T_{\mathrm{K}}(\theta, p)$, 
when $\delta\epsilon \ll T_{\mathrm{K}}$.
$T_{\mathrm{K}}(\theta, p)$ is
a decreasing function with respect to
$p\sqrt{\cos^2(\theta/2)+v^2\sin^2(\theta/2)}$.
When $\delta\epsilon$ is relevant, we evaluate
$T_{\mathrm{K}}(\delta\epsilon, \theta, p)$
using the slave-boson mean-field theory. The Kondo resonance is
split into two by finite $\delta\epsilon$, which results in the
spin accumulation in the quantum dot and
suppression of the Kondo effect.
\end{abstract}

\maketitle

\section{Introduction}

The Kondo effect, a typical many-body effect in metals
with dilute magnetic impurities, has given exciting topics
for decades.\cite{hew} The Kondo effect gives rise to the formation of
local singlet states between itinerant electrons and localized spins.
A characteristic energy scale, Kondo temperature
$T_{\mathrm{K}}$, corresponds to the binding energy of the
singlet states. At temperature $T \lesssim T_{\mathrm{K}}$,
the fluctuation of localized spins is significantly enhanced by
the spin-flip scattering of conduction electrons at the Fermi level.
Simultaneously, the scattering of the conduction electrons is
magnified, and in consequence the resistance increases in bulk metals.

Recent progress in fine processing technologies
has enabled us to make artificial atoms on semiconductors,
quantum dots (QDs). In QDs, discrete energy levels are
occupied by a fixed number of electrons by the Coulomb blockade.
With an odd number of electrons, the QDs behave like magnetic
impurities.
The Kondo effect has been observed when they are coupled to
external leads.\cite{exp1,exp2} In this case, the Kondo effect
enhances the conductance through the QDs since the conduction
electrons are resonantly transported through the Kondo singlet
state.

Lately, spin-dependent transport has been of great interest
from a viewpoint of spin-based electronics, so-called
spintronics.\cite{spintronics}
A notable example is the tunnel magnetresistance in a spin-valve
geometry, in which two ferromagnetic leads are separated by
a non-magnetic insulating layer.\cite{jul}
The tunnel current is proportional to
$\rho_{\uparrow}^2+\rho_{\downarrow}^2$ in the parallel (P)
alignment of two ferromagnets, where $\rho_{\uparrow}$ and
$\rho_{\downarrow}$ are the density of states for majority
and minority spins, respectively. It is proportional to
$2\rho_{\uparrow}\rho_{\downarrow}$ in the antiparallel (AP)
alignment, which is smaller than that in the P alignment
for $\rho_{\uparrow}>\rho_{\downarrow}$.
When the P alignment is changed to AP alignment by applying
a magnetic field, an extremely large magnetresistance is
observed.\cite{miy,moo,slo}
The spin-valve effect in a QD sandwiched by two ferromagnetic
leads has also been investigated.\cite{kon,bra}

In this article, we pay attention to the Kondo effect in the
spin-valve geometry including a QD.
This Kondo effect has been studied theoretically
\cite{ser,bul,mar1,jma,don,zha,lop,mar2,cho,uts,swi}
with motives elucidating how the ferromagnetism of
the leads influences the fluctuation of a localized spin.
One of the theoretical problems was 
whether the energy levels in the QDs are spin-split
\cite{mar1,don} or not\cite{zha,lop} in the absence of magnetic field. 
The numerical renormalization group technique, 
which is one of the most reliable methods 
to treat the Kondo physics, 
concluded the existence of the spin splitting,
$\delta\epsilon=\epsilon_{\downarrow}-\epsilon_{\uparrow}$.
\cite{mar2,cho} The spin splitting arises from the spin-dependent charge
fluctuations by the tunnel coupling to the ferromagnets and
weakens the Kondo effect.\cite{uts}
The spin splitting has been observed in experiments 
as the suppression of zero-bias anomaly 
of the differential conductance, which is a hallmark of the
Kondo effect. \cite{pas,nyg}
The splitting is the largest in the P alignment and smallest
in the AP alignment, in accordance with the theoretical results.

Until now, the Kondo effect in QD spin-valves has been elucidated
for collinear cases of two ferromagnetic leads, P or AP
alignments, by numerous papers.\cite{bul,mar1,don,zha,lop,mar2,cho,uts}
However, there exist very scarce theoretical
investigations for noncollinear alignments.\cite{ser,jma,swi}
The purpose of this article is to systematically study the Kondo effect
in QDs coupled to noncollinear ferromagnetic leads. The
spin splitting and Kondo temperature $T_{\mathrm{K}}$ are evaluated
for arbitrary alignments of the ferromagnetic leads.

In the first part of this article,
we generalize the discussion based on the scaling approach
adopted for P and AP alignments in Ref.\ \onlinecite{mar1}.
For the Kondo physics, all the energies from $T_{\mathrm{K}}$ to
the upper cutoff (band width $D_0$) have to be taken into account
properly. We use the poor man's scaling procedure\cite{and}
in two stages.\cite{hal}
First, we reduce the energy scale $D$ until the charge
fluctuation is quenched [$D=D_1$ given by Eq.\ (\ref{eqd1})]
and renormalize the energy levels $\epsilon_{\uparrow,\downarrow}$
in a QD (first stage scaling). Then we obtain the spin splitting
$\delta\epsilon$, which
is directed in an intermediate direction between the magnetic
moments in the two leads. $\delta\epsilon \propto
p\sqrt{\cos^2(\theta/2)+v^2\sin^2(\theta/2)}$ in the absence of
magnetic field, where $p$ is the spin polarization in the leads,
$\theta$ is the angle between the magnetic moments,
and $v$ is an asymmetric factor of tunnel barriers ($-1<v<1$).

We proceed to the second stage of the scaling for $D<D_1$,
where a localized spin fluctuates in the QD with fixed spin
splitting $\delta\epsilon$. The exchange couplings for the localized
spin increase with decreasing $D$ until the scaling stops at
$D=\delta\epsilon$ or reaches a fixed point of the strong
coupling limit corresponding to the Kondo effect ($D=T_{\mathrm{K}}$).
When $\delta\epsilon \ll T_{\mathrm{K}}$, the Kondo effect
takes place.
We obtain an analytical expression of Kondo temperature
$T_{\mathrm{K}}$ as a function of $\theta$ and $p$.
$T_{\mathrm{K}}(\theta, p)$ is a decreasing function with respect to
$p\sqrt{\cos^2(\theta/2)+v^2\sin^2(\theta/2)}$.
When $\delta\epsilon \gg T_{\mathrm{K}}$, no Kondo effect is
expected.

When $\delta\epsilon$ is comparable to $T_{\mathrm{K}}$, the
scaling method does not work.
In the second part of this article, we evaluate
$T_{\mathrm{K}}(\delta\epsilon, \theta, p)$
using the slave-boson mean-field (SBMF) theory.\cite{col}
The SBMF theory describes the Kondo resonant state on the
assumption of its presence and Fermi liquid behavior.
The theory is exact for the Kondo effect in
$(N=\infty)$-fold degenerate Anderson model with infinite $U$
in the case of non-magnetic leads.
For $N=2$ (spin $S=1/2$), it is widely used for the
semi-quantitative estimation of the Kondo temperature,
the Kondo resonance at the Fermi level, and resonant transmission.
However,
we have to take a special care to apply the SBMF to the present
case with magnetic leads.
In the slave-boson formalism, the empty state in the QD is
expressed by an auxiliary boson field $b$ and singly-occupied
states with spin $\sigma=\uparrow,\downarrow$ are
by pseudo-fermion operators $f_{\sigma}$. The SBMF theory
describes the Kondo effect by the condensation of the boson
field, replacing the boson operator $b$ to be a $c$ number.
We observe the formation of the Kondo resonant state
at the Fermi level with the width of $T_{\mathrm{K}}$.
In this mean-field theory, the spin fluctuation (described by
$f_{\sigma}$) is considered to construct the Kondo
resonance on one hand, the charge fluctuation is neglected
($b$ is replaced by a constant) on
the other hand. Indeed the spin splitting $\delta \epsilon$
cannot be evaluated by the SBMF theory.\cite{bul,lop}
To discuss the Kondo effect induced by the spin fluctuation,
we apply the SBMF theory to the renormalized Hamiltonian with
the band width $D_1$ after the spin splitting
$\delta \epsilon$ by the charge fluctuation has been determined
in the first stage scaling.

When $\delta\epsilon=0$, the SBMF theory yields the formation
of the Kondo resonant state at the Fermi level.
The Kondo temperature, which
is defined by the width of the resonance, decreases with increasing
$p\sqrt{\cos^2(\theta/2)+v^2\sin^2(\theta/2)}$, in semi-quantitative
agreement with that obtained by the scaling method. This
indicates an applicability of the SBMF theory to the present
problem. A finite $\delta\epsilon$ splits the Kondo resonance
into two and suppresses the Kondo effect. We evaluate
$T_{\mathrm{K}}(\delta\epsilon, \theta, p)$ by the width
of the split resonances. We also calculate the conductance $G$
at temperature $T=0$, which is
determined by the resonant tunneling through the Kondo
resonant state. $G$ could show a non-monotonic behavior as a
function of $\delta\epsilon$, in contrast to $T_{\mathrm{K}}$.
This is ascribable to an interference effect
between two spin components, analogous to the Fano resonance.

This article is organized as follows. 
In the next section (Sec.\ II), our model is presented. 
In Sec.\ III, the spin splitting $\delta\epsilon$ of the QD level 
and the Kondo temperature $T_{\mathrm{K}}$ are calculated using 
the poor man's scaling method.
In Sec.\ IV, we examine $T_{\mathrm{K}}$ and conductance 
by the SBMF theory. The conclusions are given in Sec.\ V.

\section{Model}

\begin{figure}[t]
\begin{center}
 \includegraphics[scale=0.38]{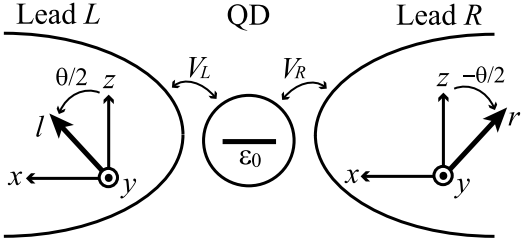}
\caption{A 
quantum dot (QD) connected to noncollinear
ferromagnetic leads, $L$ and $R$, through 
the tunnel barriers, $V_L$ and $V_R$, respectively.
The magnetic moment in lead $L(R)$ is tilted by
$\theta/2$ ($-\theta/2$)
from  the $z$ axis in the $z$-$x$ plane.
The spin polarization in the leads is given by $p$.
A single level $\epsilon_0$ is considered in the QD. }
\label{system}
\end{center}
\end{figure}

We consider a single-level QD coupled to two ferromagnetic
leads as shown in Fig.\ \ref{system}. The ferromagnets are
identical and aligned in a noncollinear way:
The magnetic moment in the lead $L$ ($R$)
is tilted by $\theta/2$ ($-\theta/2$) 
from the $z$ axis in the $z$-$x$ plane
($0 \le \theta \le \pi$).
This model is described by the Anderson Hamiltonian,
\begin{eqnarray}
H&=&\sum_{k,\sigma}\epsilon_{k\sigma}c_{Lk,l\sigma}^{\dagger}c_{Lk,l\sigma}
+\sum_{k,\sigma}\epsilon_{k\sigma}c_{Rk,r\sigma}^{\dagger}c_{Rk,r\sigma}
\nonumber\\
&&+\sum_{\sigma}\epsilon_{0,z\sigma}d_{z\sigma}^{\dagger}d_{z\sigma}+
Ud_{z\uparrow}^{\dagger}d_{z\uparrow}d_{z\downarrow}^{\dagger}
d_{z\downarrow}\nonumber\\
&&+\sum_{k,\sigma}(V_{L}c_{Lk,l\sigma}^{\dagger}d_{l\sigma}+
V_{R}c_{Rk,r\sigma}^{\dagger}d_{r\sigma}+\textrm{h.c.}),
\label{anderson}
\end{eqnarray}
where $c_{L k, l \sigma}$ ($c_{R k, r \sigma}$)
is a fermion operator of an electron
with wavenumber $k$ and spin $l \sigma$ ($r \sigma$)
in lead $L$ ($R$), and $d_{z\sigma}$ is 
that with spin $z\sigma$ in the QD level. 
We denote $l\uparrow/l\downarrow$ ($r\uparrow/r\downarrow$)
for majority/minority spin
in lead $L$ ($R$) and $z\uparrow/z\downarrow$ for spin-up/down
in the $z$ direction in the QD.
Fermion operators in the QD, $d_{l \sigma}$, $d_{r \sigma}$, 
are related to $d_{z\sigma}$ by
\begin{align}
&
\left\{
\begin{array}{l}
d_{l\uparrow}=\cos\frac{\theta}{4}d_{z\uparrow}+\sin\frac{\theta}{4}
d_{z\downarrow},\\
d_{l\downarrow}=-\sin\frac{\theta}{4}d_{z\uparrow}+\cos\frac{\theta}{4}
d_{z\downarrow},
\end{array}
\right.
\\
&
\left\{
\begin{array}{l}
d_{r\uparrow}=
\cos\frac{\theta}{4}d_{z\uparrow}-\sin\frac{\theta}{4}
d_{z\downarrow},\\
d_{r\downarrow}=\sin\frac{\theta}{4}d_{z\uparrow}+\cos\frac{\theta}{4}
d_{z\downarrow}. 
\end{array}
\right.
\end{align} 

The density of states in the ferromagnetic leads is constant,
$\rho_{\uparrow}$ and $\rho_{\downarrow}$ for majority and
minority spins, 
respectively ($\rho_{\uparrow}\geq \rho_{\downarrow}$),
in the band of $-D_0 \le \omega \le D_0$. 
The spin polarization of the leads is given by
\begin{equation}
p=\frac{\rho_{\uparrow}-\rho_{\downarrow}}
       {\rho_{\uparrow}+\rho_{\downarrow}}
\end{equation}
($0\leq p\leq 1$).

In the QD, a single energy level $\epsilon_0$ is considered.
The Zeeman energy, $E_{\mathrm{Z }}=g\mu_{{\mathrm{B}}}B$,
can be taken into account
\begin{eqnarray}
\epsilon_{0,z\uparrow/z\downarrow}=\epsilon_0\pm 
\frac{E_{\mathrm{Z}}}{2},
\end{eqnarray}
where $B$ is an external magnetic field in the $z$ direction.
The magnetic field produced by the ferromagnetic leads may be
included in $B$.
The Coulomb interaction for double occupancy in the QD, $U$,
is assumed to be infinitely large.

The strength of tunnel coupling to the leads is characterized
by
\begin{eqnarray}
\Delta=\pi (\rho_{\uparrow}+\rho_{\downarrow})
(|V_{L}|^2+|V_{R}|^2).
\label{tunnel-strength}
\end{eqnarray}
We define an asymmetric factor of tunnel couplings
by
\begin{eqnarray}
v=\frac{V_{L}^2-V_{R}^2}{V_{L}^2+V_{R}^2}
\label{asymmetric-f}
\end{eqnarray}
($-1<v<1$). In the text, we treat the case of symmetric barriers,
$V_{L}=V_{R}\equiv V$ ($v=0$). Then we can 
regard the $z$ axis
as a well-defined quantization axis in the QD from the symmetry of
the system. The generalization to the case of $V_{L} \not= V_{R}$
($v \not=0$) is straight-forward, as explained in Appendix A.

To extract conduction modes which couple to the QD,
we perform a unitary transformation for the conduction electrons
in the two leads. In the energy range of $[\epsilon, \epsilon+d\epsilon]$,
the number of majority spins, $\rho_{\uparrow}d\epsilon$, is
larger than that of minority spins, $\rho_{\downarrow}d\epsilon$,
by $(\rho_{\uparrow}-\rho_{\downarrow})d\epsilon$ in each lead.
$\rho_{\downarrow}d\epsilon$ of majority spins are paired with
minority spins,
\begin{align}
\left\{
\begin{array}{l}
a_{k\uparrow}=\cos\frac{\theta}{4}c_{k\uparrow}^{(s)}
-\sin\frac{\theta}{4}c_{k'\downarrow}^{(a)},\\
\bar{a}_{k\uparrow}=\cos\frac{\theta}{4}c_{k\uparrow}^{(a)}
+\sin\frac{\theta}{4}c_{k'\downarrow}^{(s)},\\
a_{k\downarrow}=\cos\frac{\theta}{4}c_{k'\downarrow}^{(s)}
-\sin\frac{\theta}{4}c_{k\uparrow}^{(a)},\\
\bar{a}_{k\downarrow}=\cos\frac{\theta}{4}c_{k'\downarrow}^{(a)}
+\sin\frac{\theta}{4}c_{k\uparrow}^{(s)},
\end{array}
\right.
\label{trans-a1}
\end{align}
where $\epsilon \le \epsilon_{k\uparrow}, \epsilon_{k'\downarrow}
< \epsilon+d\epsilon$.
The rest of majority spins is transformed by
\begin{align}
\left\{
\begin{array}{l}
a_{k\uparrow}=c_{k\uparrow}^{(s)},\\
a_{k\downarrow}=-c_{k\uparrow}^{(a)}.
\end{array}
\right.
\label{trans-a2}
\end{align}
Here, we have introduced the symmetric and anti-symmetric combinations
for the conduction electrons in the two leads,
\begin{align}
\left\{
\begin{array}{l}
c_{k\uparrow}^{(s)}=\frac{1}{\sqrt{2}}\left(c_{Lk,l\uparrow}
+c_{Rk,r\uparrow}\right),\\
c_{k\uparrow}^{(a)}=\frac{1}{\sqrt{2}}\left(-c_{Lk,l\uparrow}
+c_{Rk,r\uparrow}\right),\\
c_{k\downarrow}^{(s)}=\frac{1}{\sqrt{2}}\left(c_{Lk,l\downarrow}+
c_{Rk,r\downarrow}\right),\\
c_{k\downarrow}^{(a)}=\frac{1}{\sqrt{2}}\left(c_{Lk,l\downarrow}
-c_{Rk,r\downarrow}\right).
\end{array}
\right.
\label{ct}
\end{align}
Note that index $\sigma=\uparrow,\downarrow$ of 
$a_{k\sigma}$ and $\bar{a}_{k\sigma}$
does not any more mean majority or minority spins.
By this transformation, the tunnel Hamiltonian $H_{\mathrm{T }}$
[the last term in Eq.\ (\ref{anderson})] is rewritten as
\begin{align}
H_{\mathrm{T}} = &
\sqrt{2}V\sum_{k}{}'\sum_{\sigma}(
a_{k\sigma}^{\dagger}d_{z\sigma}+\textrm{h.c.})
\nonumber\\
+&
 \sqrt{2}V\sum_{k}{}'' \left(
\cos\frac{\theta}{4}a_{k\uparrow}^{\dagger}d_{z\uparrow}+
\sin\frac{\theta}{4}a_{k\downarrow}^{\dagger}d_{z\downarrow}
+\textrm{h.c.} \right).
\end{align}
The summations of $\sum'$ and $\sum''$ correspond to the density
of states $\rho_{\downarrow}$ and $\rho_{\uparrow}-\rho_{\downarrow}
\equiv \Delta \rho$, respectively.
The modes $\bar{a}_{k\uparrow}$ and $\bar{a}_{k\downarrow}$ are
completely decoupled from the QD and hence they can be
disregarded in the discussion of spin splitting of the QD level and
Kondo effect.

The total Hamiltonian (\ref{anderson}) is rewritten as
\begin{align}
H=& \sum_{k,\sigma}\epsilon_{k\sigma}
a_{k\sigma}^{\dagger}a_{k\sigma}
\nonumber\\
+& \sum_{\sigma}\epsilon_{0,z\sigma}d_{z\sigma}^{\dagger}d_{z\sigma}
+Ud_{z\uparrow}^{\dagger}d_{z\uparrow}
d_{z\downarrow}^{\dagger}d_{z\downarrow}
\nonumber\\
+& \sqrt{2}V
\sum_{\sigma}(A_{\sigma}^{\dagger}d_{z\sigma}+\textrm{h.c.}),
\label{anderson2}
\end{align}
where
\begin{eqnarray}
A_{\uparrow} &=& \sum_{k}{}'a_{k\uparrow}
+\cos\frac{\theta}{4}\sum_{k}{}''a_{k\uparrow},
\label{A-up}
\\
A_{\downarrow} &=& \sum_{k}{}'a_{k\downarrow}
+\sin\frac{\theta}{4}\sum_{k}{}''a_{k\downarrow}.
\label{A-down}
\end{eqnarray}
The summation over $k$ is $\sum=\sum'+\sum''$
in the first term in Hamiltonian (\ref{anderson2}).
The corresponding density of states is $\rho_{\uparrow}$.

\section{Scaling Theory}

In this section, we apply the poor man's scaling theory\cite{and}
to the Hamiltonian (\ref{anderson2}) and evaluate the
spin splitting of QD level $\delta\epsilon$
and Kondo temperature.
The scaling procedure consists of two stages.\cite{hal}
In the first stage (Sec.\ III.\ A),
we reduce the energy scale from the band width $D_0$ to
$D_1$ where the charge fluctuation is
quenched. 
By integrating out the excitations in the energy range of
$D_1<D<D_0$, we renormalize the energy levels in the QD,
$\epsilon_{\uparrow}$ and $\epsilon_{\downarrow}$, and
determine the spin splitting
$\delta\epsilon=\epsilon_{\downarrow}-\epsilon_{\uparrow}$.
Even when $B=0$, we find a finite $\delta\epsilon$.
In Sec.\ III.\ B, we proceed to the second stage scaling for
$D<D_1$ where the spin fluctuation is dominant.
We make the Kondo Hamiltonian and discuss the Kondo effect.

\subsection{Spin splitting of QD level}

We examine the Coulomb blockade region with one electron in the QD,
$-D_0 \ll \epsilon_0 \ll \mu-\Delta$
and $\mu+\Delta \ll \epsilon_0+U$,
where $\mu \equiv 0$ is the Fermi energy of conduction electrons
in the leads.
The Coulomb interaction $U$ is assumed to be strong enough
to forbid double occupancy in the QD; $U\to\infty$.

The number of electrons in the QD fluctuates
between zero and one through $H_{\mathrm{T}}$.
We denote $\left| 0\right\rangle$ for the empty state and
$\left| z\sigma\right\rangle =d_{z\sigma}^{\dagger}\left|
0\right\rangle$ for singly occupied state with spin $z\sigma$.
They have the energies $E_0$ and $E_{z\sigma}$,
respectively.
Initially, $E_0=0$ and $E_{z\sigma}=\epsilon_{0,z\sigma}$.

Under the instruction of the poor man's scaling,  
we renormalize the energies $E_0$ and $E_{z\sigma}$ by
integrating out the high energy excitations in the conduction
band. Reducing half of the band width from $D$ to $D-|dD|$, 
$E_0$ and $E_{z\sigma}$ are renormalized as 
$E_0+dE_0$ and $E_{z\sigma}+dE_{z\sigma}$, where
\begin{align}
dE_0=
&-\frac{2|V|^2|dD|}{D+E_{z\uparrow}-E_0}
\left( \rho_{\downarrow}+\Delta\rho \cos^2\frac{\theta}{4} \right)
\nonumber\\
&-\frac{2|V|^2|dD|}{D+E_{z\downarrow}-E_0}
\left( \rho_{\downarrow}+\Delta\rho \sin^2\frac{\theta}{4} \right),
\\
dE_{z\uparrow}=
&-\frac{2|V|^2|dD|}{D+E_{0}-E_{z\uparrow}}
\left( \rho_{\downarrow}+\Delta\rho \cos^2\frac{\theta}{4} \right),
\\
dE_{z\downarrow}=
&-\frac{2|V|^2|dD|}{D+E_{0}-E_{z\downarrow}}
\left( \rho_{\downarrow}+\Delta\rho \sin^2\frac{\theta}{4} \right),
\end{align}
within the second-order perturbation with respect to
$H_{\mathrm{T }}$.
For $D\gg |E_{z\sigma}-E_0|$, we obtain the
spin-dependent scaling equations for the renormalized energy
levels,
$\epsilon_{z\sigma}=E_{z\sigma}-E_0$,
\begin{eqnarray}
\frac{d\epsilon_{z\sigma}}{d\ln D}=
-\frac{\Gamma_{\bar{\sigma}}}{\pi},
\label{ScalingEq1}
\end{eqnarray}
where
\begin{eqnarray}
\Gamma_{\uparrow/\downarrow}=\frac{\Delta}{2}
\left( 1\pm p\cos\frac{\theta}{2} \right)
\end{eqnarray}
is the level broadening for spin
$z\uparrow/z\downarrow$ in the QD.
Note that the renormalization of $\epsilon_{z\sigma}$
depends on the coupling strength for the opposite spin,
$\Gamma_{\bar{\sigma}}$.

Integrating Eq.\ (\ref{ScalingEq1}) from 
($D_0,\epsilon_{0,z\sigma}$) 
to ($D_1,\epsilon_{z\sigma}$),
we obtain
\begin{eqnarray}
 \epsilon_{z\sigma}=\epsilon_{0,z\sigma}
+\frac{\Gamma_{\bar{\sigma}}}{\pi}\ln\frac{D_0}{D_1},
\label{re-levels}
\end{eqnarray}
where $D_1 \approx -\epsilon_{z\uparrow}$ at which 
the perturbation theory breaks down. \cite{hal}
$D_1$ satisfies the equation
\begin{eqnarray}
-D_1\approx
\epsilon_{0,z\uparrow}+\frac{\Gamma_{\downarrow}}{\pi} \ln \frac{D_0}{D_1}.
\label{eqd1}
\end{eqnarray} 
The charge fluctuation is quenched at the energy scale of
$D \sim D_1$.
As a result,
the renormalized spin splitting is given by
\begin{eqnarray}
\delta\epsilon \equiv\epsilon_{z\downarrow}-\epsilon_{z\uparrow}
=\frac{\Delta}{\pi}p\cos\frac{\theta}{2}\ln\frac{D_0}{D_1}-E_{\mathrm{Z }}.
\label{split1}
\end{eqnarray}
This has been obtained for the P alignment ($\theta=0$)
in Ref.\ \onlinecite{mar1}.

\begin{figure}[t]
\begin{center}
 \rotatebox{-90}{\includegraphics[scale=0.55]{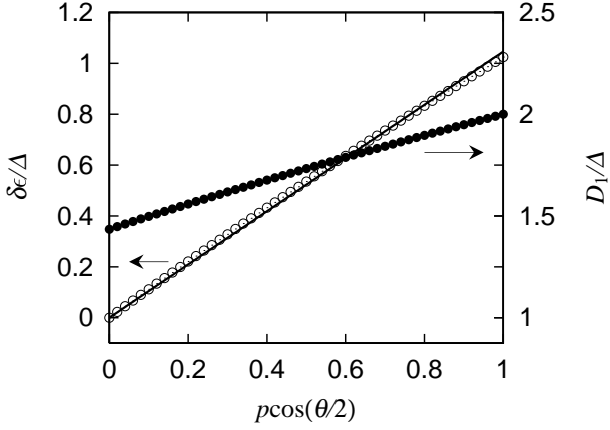}}
\caption{Spin splitting
$\delta\epsilon=\epsilon_{z\downarrow}-\epsilon_{z\uparrow}$  
of the QD level (open circles) as a function of
$p\cos(\theta/2)$, when
$D_0/\Delta=50$, $\epsilon_0/\Delta=-2$, and $E_{\mathrm{Z }}=0$.
The tunnel barriers are symmetric ($v=0$).
$\delta\epsilon$ is fitted to a straight line,
$\delta\epsilon/\Delta=1.046p\cos(\theta/2)$ (solid line).
Closed circles indicate $D_1$ which is determined by
Eq.\ (\ref{eqd1}).
}
\label{delep}
\end{center}
\end{figure}

We observe a finite $\delta\epsilon$ in the absence of magnetic
field, $E_{\mathrm{Z }}=0$. We estimate the value in the experimental
situation using $\mathrm{C}_{60}$,\cite{pas}
where $D_0/\Delta=50$ and $\epsilon_0/\Delta=-2$.
First, $D_1$ is determined by solving Eq.\ (\ref{eqd1}). Then
Eq.\ (\ref{split1}) with $E_{\mathrm{Z }}=0$ yields the spin
splitting $\delta\epsilon$.
Figure \ref{delep} shows $\delta\epsilon$ and $D_1$ as functions of
$p\cos(\theta/2)$. Since $D_1$ depends on $p\cos(\theta/2)$
only weakly, $\delta\epsilon$ is almost linearly proportional 
to $p\cos(\theta/2)$ (the fitting is indicated by solid line
in Fig.\ 2).
In the presence of spin polarization, $p\not=0$,
the spin splitting decreases with increasing $\theta$ from $0$ to $\pi$.
$\delta\epsilon$ is maximal in the P alignment ($\theta=0$) and zero
in the AP alignment ($\theta=\pi$).
In the experimental result,\cite{pas} $\delta\epsilon=9\mathrm{meV}$
with $p=0.31$ and $\Delta=30\mathrm{meV}$, and hence
$\delta\epsilon/\Delta=0.33$ in the P alignment.
The calculated result indicates $\delta\epsilon/\Delta=0.338$ when
$p=0.31$ and $\theta=0$, which is in good agreement with the
experimental observation.

In the case of asymmetric barriers, $V_L\not= V_R$, the spin
splitting is calculated in Appendix A [see Eq.\ (\ref{g-delep})],
\begin{eqnarray}
\delta\epsilon=\frac{\Delta}{\pi}
p\sqrt{\cos^2\frac{\theta}{2}+v^2\sin^2\frac{\theta}{2}}
\ln\frac{D_0}{D_1}
-E_{\mathrm{Z }},
\label{g-delep0}
\end{eqnarray}
with an asymmetric factor $v$ in Eq.\
(\ref{asymmetric-f}).\cite{com5}
In the absence of magnetic field
($E_{\mathrm{Z}}=0$), $\delta\epsilon$ is always maximal at
$\theta=0$ and minimal at $\theta = \pi$ although
$\delta\epsilon \ne 0$ in the AP alignment
($\theta = \pi$) in general.
The observation of a finite spin-splitting with AP
alignment\cite{pas} may be explained by
Eq.\ (\ref{g-delep0}) with asymmetric barriers ($v \not= 0$).

\subsection{Kondo effect}

In the energy scale of $D \ll D_1$, the empty state
$\left| 0\right\rangle$ is irrelevant:
The number of electrons in the
QD is fixed to be $\left\langle n_d \right\rangle\simeq 1$.
To restrict the QD states to 
$\left| z \uparrow \right\rangle$ or
$\left| z \downarrow \right\rangle$, we apply the Schrieffer-Wolff
transformation to the Anderson model with renormalized levels,
$\epsilon_{z\uparrow}$ and $\epsilon_{z\downarrow}$, given
by Eq.\ (\ref{re-levels}).
We obtain the Kondo Hamiltonian
\begin{align}
H_{\mathrm{Kondo}} =& H_0+H_{\mathrm{sd}},
\label{KondoH}
\\
H_0=&
\sum_{k,\sigma}\epsilon_{k\sigma}
a_{k\sigma}^{\dagger}a_{k\sigma}
+\sum_{\sigma}\epsilon_{z\sigma} d_{z\sigma}^{\dagger}d_{z\sigma},
\\
H_{\mathrm{sd}}=& J_{+}S^{+}
A_{\downarrow}^{\dagger}A_{\uparrow}+
J_{-}S^{-}A_{\uparrow}^{\dagger}A_{\downarrow}
\nonumber\\
&+S^{z}(J_{z\uparrow}A_{\uparrow}^{\dagger}A_{\uparrow}
-J_{z\downarrow}A_{\downarrow}^{\dagger}A_{\downarrow}),
\label{sd}
\end{align}
where $S$'s are spin operators in the QD,
$S^{+}=d_{z\uparrow}^{\dagger}d_{z\downarrow}$,
$S^{-}=d_{z\downarrow}^{\dagger}d_{z\uparrow}$, and
$S^{z}=(d_{z\uparrow}^{\dagger}d_{z\uparrow}
-d_{z\downarrow}^{\dagger}d_{z\downarrow})/2$.
Here, we have neglected a small
level shift, $-D_{1}
[(\rho_{\uparrow}+\rho_{\downarrow})/2]
\sum_{\sigma}J_{z\sigma}[1+\sigma p\cos(\theta/2)]
d_{z\sigma}^{\dagger}d_{z\sigma}$,
and potential scattering terms: The former may be
included in the second term in $H_0$, whereas the latter
is not relevant to the Kondo effect.
$J_+=J_-\equiv J_{\perp}$ and $J_{z\sigma}$ are
exchange coupling constants.
They are given by
\[
J_{\perp}=V^2 \left( \frac{1}{|\epsilon_{z\uparrow}|}+
\frac{1}{|\epsilon_{z\downarrow}|} \right), \ \ \ 
J_{z\sigma}=\frac{2V^2}{|\epsilon_{z\sigma}|}.
\]

We continue the scaling procedure to $D < D_1$, now using the 
Kondo Hamiltonian (\ref{KondoH}).
We obtain the scaling equations 
for $J_{\perp}$ and $J_{z\sigma}$ to the second-order with respect
to $H_{\mathrm{sd}}$ when $D \gg \delta \epsilon$.
\begin{align}
&\frac{dk_{\perp}}{d\ln D}=-(k_{\uparrow}+
k_{\downarrow})k_{\perp},
\label{ScalingEq2a} \\
&\frac{dk_{\sigma}}{d\ln D}=-2k_{\perp}^2, 
\label{ScalingEq2b}
\end{align}
where $k_{\perp}$ and $k_{\sigma}$ are
effective coupling constants,
$k_{\perp}=
[(\rho_{\uparrow}+\rho_{\downarrow})/2]J_{\perp}
\sqrt{1-p^2\cos^2(\theta/2)}$ and
$k_{\sigma}=
[(\rho_{\uparrow}+\rho_{\downarrow})/2]J_{z\sigma}
\left[1+\sigma p\cos(\theta/2)\right]$.

When $D \ll \delta \epsilon$,
state $\left| z \downarrow \right\rangle$ is irrelevant and
hence the spin-flip processes do not take place. The
coupling constants cease to increase with decreasing $D$.

Let us consider a situation where $\delta\epsilon$ is
negligibly small ($\delta\epsilon \ll T_{\mathrm{K}}$).
This situation can be realized by tuning
an external magnetic field.\cite{mar1} Then the
scaling equations, (\ref{ScalingEq2a}) and (\ref{ScalingEq2b}),
are applicable until the scaling reaches a fixed point of
strong coupling limit. As a result,
the Kondo effect takes place. 
The initial condition for the scaling is given by
\begin{equation}
J_{\perp}=J_{z\uparrow}=J_{z\downarrow} =
\frac{2V^2}{|\epsilon| } \equiv J
\end{equation}
with $D=D_1$, where $\epsilon \equiv \epsilon_{z\uparrow}
=\epsilon_{z\downarrow}$. $\epsilon=-D_1$ from the definition
of $D_1$ in the previous subsection.

From Eqs.\
(\ref{ScalingEq2a}) and (\ref{ScalingEq2b}),	
we find the scaling trajectory,
$(k_{\uparrow}
+k_{\downarrow})^2-4k_{\perp}^2
=(\rho_{\uparrow}+\rho_{\downarrow})^2
J^2p^2\cos^2(\theta/2)$ and 
$k_{\uparrow}-k_{\downarrow}
=(\rho_{\uparrow}+\rho_{\downarrow})Jp\cos(\theta/2)$.
On the trajectory, the coupling constants go to the fixed
point of
$k_{\perp}$, $k_{\uparrow}$,
$k_{\downarrow}= \infty$, with decreasing the
energy scale $D$. The energy scale where the fixed point is
reached yields the Kondo temperature,
\begin{eqnarray}
T_{\mathrm{K}}(\theta,p)
=D_1\exp\left[-\frac{1}{(\rho_{\uparrow}+
\rho_{\downarrow})J}\frac{\textrm{arctanh}
\left(p\cos\frac{\theta}{2}\right)}
{p\cos\frac{\theta}{2}}\right]\!\!.
\label{KondoTemp}
\end{eqnarray}
This is an extension of the result in Ref.\ \onlinecite{mar1}
for the P alignment.\cite{com3}

Equation (\ref{KondoTemp})
describes the dependence of the Kondo temperature on the
spin polarization $p$ and relative angle $\theta$ between the
magnetic moments in the ferromagnetic leads.
$T_{\mathrm{K}}(\theta,p)$ is a function of
$p\cos(\theta/2)$.
With an increase in $p\cos(\theta/2)$,
$T_{\mathrm{K }}$ decreases as shown in Fig.\ 3 (solid line),
reflecting the suppression of spin fluctuation in the QD.
When the leads are completely polarized ($p=1$) and 
in the P alignment ($\theta=0$), the Kondo effect vanishes
owing to the absence of spin fluctuation.
In the AP alignment ($\theta=\pi$), $T_{\mathrm{K }}$ is
independent of the spin polarization $p$ and given by
$T_{\mathrm{K}}=D_1
\exp \{-1/[(\rho_{\uparrow}+\rho_{\downarrow})J]\}$.
This expression coincides with that for non-magnetic leads
($p=0$).\cite{mar1}

\begin{figure}[t]
\begin{center}
 \rotatebox{-90}{\includegraphics[scale=0.5]{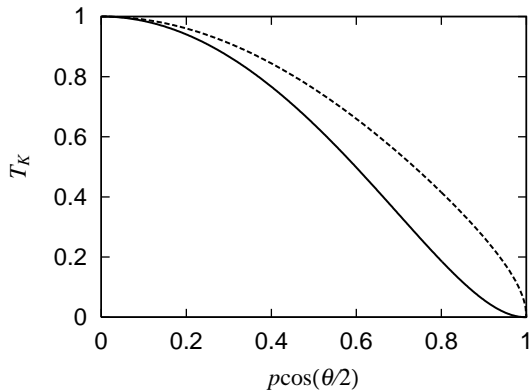}}
\caption{The Kondo temperature $T_{\mathrm{K }}$ 
obtained by the scaling calculation (solid line) and slave-boson
mean-field theory (dashed line),
as a function of $p\cos(\theta/2)$.
The tunnel barriers are symmetric ($v=0$).
The spin splitting is $\delta \epsilon=0$.
$T_{\mathrm{K }}$ is normalized by the value
at $p\cos(\theta/2)=0$.
In the scaling calculation, we choose
$(\rho_{\uparrow}+\rho_{\downarrow})J=\Delta/(\pi |\epsilon|)$
with renormalized QD level
$\epsilon=-D_1=-1.43\Delta$.\cite{com3}
}
\label{tktk}
\end{center}
\end{figure}

The properties of this Kondo effect are determined by
the fixed point of strong coupling limit.
Although $J_{\perp}$, $J_{z\uparrow}$,
and $J_{z\downarrow}\to \infty$ for
$D\to T_{\scriptsize{\textrm{K}}}$, their ratios are finite,
\begin{eqnarray*}
\frac{J_{z\uparrow}}{J_{z\downarrow}}=
\frac{1-p\cos\frac{\theta}{2}}{1+p\cos\frac{\theta}{2}},
\ \ \ 
\frac{J_{\perp}}{\sqrt{J_{z\uparrow}
J_{z\downarrow}}}= 1.
\end{eqnarray*}
The former expresses an important 
characteristic of the spin asymmetry by the coupling to
noncollinearly spin-polarized leads. The ratio of
$J_{z\uparrow}/J_{z\downarrow}$ is equal to
that of effective density of states of conduction electrons
coupled to $\left| z \downarrow \right\rangle$ and
$\left| z \uparrow \right\rangle$,
$\rho_{\downarrow}^{\ast}/\rho_{\uparrow}^{\ast}$,
where
$\rho_{\downarrow}^{\ast}=
\rho_{\downarrow}+\Delta\rho \sin^2 (\theta/4)$
and
$\rho_{\uparrow}^{\ast}=
\rho_{\downarrow}+\Delta\rho \cos^2 (\theta/4)$
[see Eqs.\ (\ref{A-up}) and (\ref{A-down})].
In other words, the relation of
$\rho_{z\uparrow}^{\ast}J_{z\uparrow}
=\rho_{z\downarrow}^{\ast}J_{z\downarrow}$ is realized
in the strong coupling limit.
In the formation of Kondo singlet state, spin state
$\left| z \uparrow \right\rangle$ in the
QD is coupled to conduction electrons $A_{\downarrow}$
with density of states $\rho_{z\downarrow}^{\ast}$ by
$J_{z\downarrow}$, whereas spin state
$\left| z \downarrow \right\rangle$ is coupled to conduction
electrons $A_{\uparrow}$ with density of states
$\rho_{z\uparrow}^{\ast}$ by $J_{z\uparrow}$.
The relation of $\rho_{z\uparrow}^{\ast}J_{z\uparrow}
=\rho_{z\downarrow}^{\ast}J_{z\downarrow}$
indicates that the spin-up and -down in the
QD are equivalently screened and a local singlet state is
realized in spite of the spin polarization in the ferromagnetic
leads.

In the case of asymmetric barriers, $V_L\not= V_R$, the Kondo
temperature $T_{\mathrm{K}}(\theta,p)$ is given by Eq.\ 
(\ref{KondoTemp2}) in Appendix A. The above-mentioned discussion
is applicable to the case of $v \not= 0$ if
$p\cos(\theta/2)$ is replaced by
$p\sqrt{\cos^2(\theta/2)+v^2\sin^2(\theta/2)}$.

We have observed the Kondo effect for
$\delta \epsilon \ll T_{\mathrm{K}}$.
In the opposite situation of $\delta \epsilon \gg T_{\mathrm{K}}$,
the scaling by Eqs.\ (\ref{ScalingEq2a}) and (\ref{ScalingEq2b})
ends at $D \sim \delta \epsilon$ and then the coupling constants
saturate. No Kondo effect is expected in this case.

\section{Slave-Boson Mean-Field Theory}

From the scaling calculation, we obtain an analytic expression
of the Kondo temperature, Eq.\ (\ref{KondoTemp}), when
the spin splitting $\delta \epsilon \ll T_{\mathrm{K}}$.
The Kondo effect does not take place when
$\delta \epsilon \gg T_{\mathrm{K}}$.
To evaluate the Kondo
temperature with $\delta \epsilon \sim T_{\mathrm{K}}$, we adopt the
slave-boson mean-field (SBMF) theory\cite{col} in this section,
assuming that $U=\infty$.
The conductance is also calculated at temperature $T=0$.

In the slave-boson formalism, we introduce an auxiliary boson
field to represent an empty state in the QD.
Using the creation operator of the boson $b^\dagger$, the empty
state in the QD is written as
$|0 \rangle=b^\dagger |\mathrm{vac} \rangle$, where
$|\mathrm{vac} \rangle$ is the vacuum state. The singly-occupied
states are expressed as
$|z\sigma \rangle=f_{z\sigma}^\dagger|\mathrm{vac} \rangle$ with
$\sigma=\uparrow, \downarrow$, where
$f_{z\sigma}^\dagger$ is a pseudo-fermion operator to create
the spin state. When $U=\infty$, the constraint
\begin{eqnarray}
b^{\dagger}b+\sum_{\sigma=\uparrow,\downarrow}
f_{z\sigma}^{\dagger}f_{z\sigma}=1
\label{constraint}
\end{eqnarray}
is required since the QD state should be empty or singly-occupied
with either spin. The original operators for an electron
in the QD are rewritten as
$d_{z\sigma}^\dagger=f_{z\sigma}^\dagger b$ and
$d_{z\sigma}=f_{z\sigma} b^\dagger$.

We apply this formalism to the Anderson model (\ref{anderson})
with renormalized QD levels, Eq.\ (\ref{re-levels}),
obtained by the first stage scaling in Sec.\ III.\ A.
Hence the charge fluctuation in the energy range of $D_1<D<D_0$ has
been taken into account.
The density of states is constant, $\rho_{\uparrow/\downarrow}$,
in the band of $-D_1 \le \omega \le D_1$.
The Hamiltonian is rewritten as
\begin{align}
H&=\sum_{k\sigma}\epsilon_{k\sigma}c_{Lk,l\sigma}^{\dagger}
c_{Lk,l\sigma}+\sum_{k\sigma}\epsilon_{k\sigma}
c_{Rk,r\sigma}^{\dagger}c_{Rk,r\sigma}\nonumber\\
&+\sum_{\sigma}\epsilon_{z\sigma}f_{z\sigma}^{\dagger}f_{z\sigma}
+\lambda \left( b^{\dagger}b+\sum_{\sigma}
f_{z\sigma}^{\dagger}f_{z\sigma}-1 \right)
\nonumber\\
&+\sum_{k\sigma}(Vc_{Lk,l\sigma}^{\dagger}
f_{l\sigma}b^{\dagger}+Vc_{Rk,r\sigma}^{\dagger}f_{r\sigma}
b^{\dagger}+\textrm{h.c.}),
\end{align}
where $\lambda$ is a Lagrange multiplier
to consider the constraint, Eq.\ (\ref{constraint}). 
In the following discussion, we fix $D_1$ and
$\epsilon \equiv (\epsilon_{z\uparrow}+\epsilon_{z\downarrow})/2$.

In the SBMF theory, the Kondo effect takes place when the
boson field is condensed: The boson operators $b$,
$b^{\dagger}$ are replaced by a $c$ number,
$\left\langle b\right\rangle\equiv r$. Then the
Hamiltonian is reduced to a non-interacting Anderson model
in which the QD levels are given by
$\tilde{E}_{z\uparrow/z\downarrow}=
\epsilon_{z\uparrow/z\downarrow}+\lambda$
and the tunnel coupling is $\tilde{V}=rV$.\cite{com4}
We can easily calculate the Green functions in the QD
\begin{align}
&\hat{G}_{z\sigma z\sigma}(\omega)
=\frac{1}{\omega-\tilde{E}_{z\sigma}
+\mathrm{i}\tilde{\Gamma}_{\sigma}},\\
&\hat{G}_{z\uparrow z\downarrow}(\omega)
=\hat{G}_{z\downarrow z\uparrow}(\omega)=0,
\end{align}
where $\tilde{\Gamma}_{\sigma}=r^2\Gamma_{\sigma}$. 
The fact that the $z$ axis is a well-defined quantization axis 
results in zero off-diagonal elements. 
Minimizing the expectation value of the Hamiltonian 
with respect to $\lambda$ and $r$, self-consistent equations 
for them are obtained as
\begin{align}
&\sum_{\sigma}\frac{\Gamma_{\sigma}}{\pi}
\ln\frac{\sqrt{\tilde{E}_{z\sigma}^2
+\tilde{\Gamma}_{\sigma}^2}}{D_1}+\lambda=0,
\label{scf-eq1} \\
&r^2+\left\langle n_f\right\rangle=1,
\label{scf-eq2}
\end{align}
where 
$\left\langle n_f\right\rangle
=\langle n_{z\uparrow}\rangle+\langle n_{z\downarrow}\rangle$,
with
\begin{eqnarray}
\langle n_{z\sigma}\rangle=
\langle f_{z\sigma}^{\dagger} f_{z\sigma}\rangle
=\frac{1}{\pi}
\arctan\frac{\tilde{\Gamma}_{\sigma}}{\tilde{E}_{z\sigma}}.
\label{nd}
\end{eqnarray}

We are interested in the case of
$\epsilon_{z\sigma} \ll \mu-\Delta$,
where $\left\langle n_f\right\rangle\simeq 1$
(Kondo regime).
By solving Eqs.\ (\ref{scf-eq1}) and (\ref{scf-eq2}),
we observe the spin-split Kondo resonant levels, which are
located at
\begin{eqnarray}
\tilde{E}_{z\uparrow/z\downarrow}=
\mp\frac{\delta\epsilon}{2}\left( 1 \pm p\cos\frac{\theta}{2}
\right),
\label{Kondolevel}
\end{eqnarray}
below and above the Fermi level $\mu=0$.
Note that they are not equi-distant from $\mu=0$
except in the AP alignment ($\theta =\pi$).
The resonant widths are given by
$\tilde{\Gamma}_{\uparrow/\downarrow}$.
We define the Kondo temperature
by the geometric mean of the resonant widths,
$T_{\mathrm{K }}=
\sqrt{\tilde{\Gamma}_{\uparrow}\tilde{\Gamma}_{\downarrow}}$.
It is given by
\begin{align}
T_{\mathrm{K}}(\delta\epsilon,\theta,p) =
\sqrt{T_{\mathrm{K}}(0,\theta,p)^2
-|\tilde{E}_{z\uparrow}\tilde{E}_{z\downarrow}|},
\label{TKd}
\end{align}
and 
\begin{align}
T_{\mathrm{K }}(0,\theta,p)=
D_1\exp\left[-\frac{\pi|\epsilon|}{\Delta}-p\cos
\frac{\theta}{2}\textrm{arctanh}\left(p\cos\frac{\theta}{2}\right)\right].
\label{Tk0}
\end{align}
In Eq.\ (\ref{Tk0}), $\delta\epsilon=0$ and thus $\epsilon=
\epsilon_{z\uparrow}=\epsilon_{z\downarrow}$ ($=-D_1$).

\begin{figure}[t]
\begin{center}
 \rotatebox{-90}{\includegraphics[scale=0.34]{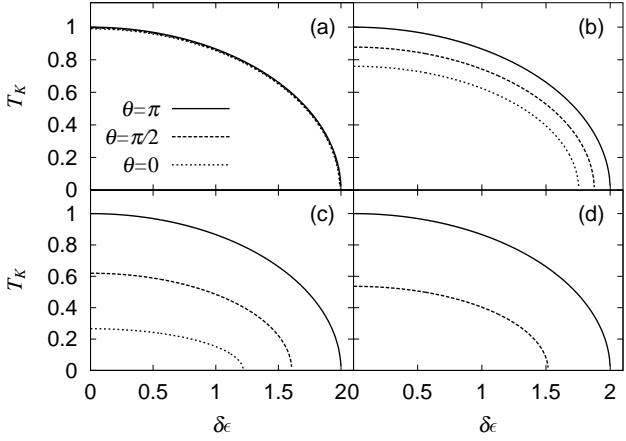}}
\caption{The Kondo temperature $T_{\mathrm{K }}$ obtained
by the slave-boson mean-field theory as
a function of the spin splitting $\delta\epsilon$.
The tunnel barriers are symmetric ($v=0$).
The spin polarization in the ferromagnetic leads
is (a) $p=0.1$, (b) $0.5$, (c) $0.9$, and (d) $1$.
The angle $\theta$ between the magnetic moments in the leads
is $0$, $\pi/2$, and $\pi$.
Both $T_{\mathrm{K }}$ and $\delta\epsilon$ are normalized by
$T_{\mathrm{K }}(\delta\epsilon=0, \theta=\pi,p)$. Note that
$T_{\mathrm{K }}=0$ for any $\delta\epsilon$ in the case
of (d) $p=1$ and $\theta=0$.
}
\label{Tk}
\end{center}
\end{figure}

First, we compare Eq.\ (\ref{Tk0}) with Eq.\ (\ref{KondoTemp})
obtained by the scaling theory when $\delta\epsilon=0$.
In the case of non-magnetic leads ($p=0$) 
or of magnetic leads in the AP alignment ($\theta=\pi$),
they are identical to each other:
$T_{\mathrm{K }}=D_1\exp(-\pi|\epsilon|/\Delta)
=D_1 \exp \{-1/[(\rho_{\uparrow}+\rho_{\downarrow})J]\}$.
In the case of magnetic leads, both of them are
functions of $p\cos(\theta/2)$. They decrease with increasing
$p\cos(\theta/2)$ similarly, as shown in Fig.\ \ref{tktk}.
The Kondo effect is the strongest at $p\cos(\theta/2)=0$,
whereas it disappears $p\cos(\theta/2)=1$.
This semi-quantitative agreement clearly indicates an
applicability of the SBMF theory to the present problem.

Second, we discuss the Kondo effect with finite $\delta\epsilon$.
In Fig.\ \ref{Tk}, we plot
$T_{\mathrm{K }}(\delta\epsilon,\theta,p)$ in
Eq.\ (\ref{TKd}) as a function of $\delta\epsilon$.
It decreases monotonically with an increase in $\delta\epsilon$,
reflecting the separation between the Kondo resonant 
levels for $z\uparrow$ and $z\downarrow$ in Eq.\
(\ref{Kondolevel}).
The Kondo temperature vanishes at $\delta\epsilon=
2T_{\mathrm{K}}(0,\theta,p)/\sqrt{1-p^2\cos^2(\theta/2)}$.
Note that this sudden disappearance of the Kondo
effect is an artifact by the SBMF theory. The theory works
unless the Kondo temperature is too small.\cite{com6}

The suppression of the Kondo effect can be also understood
in terms of the spin accumulation,
$\left\langle \delta n_f\right\rangle =
\left\langle n_{z\uparrow}\right\rangle-
\left\langle n_{z\downarrow} \right\rangle$.
Using Eq.\ (\ref{nd}) and $\left\langle n_f\right\rangle \simeq 1$, 
the Kondo temperature
is rewritten as
\begin{align}
T_{\mathrm{K}}(\delta\epsilon)&\simeq T_{\mathrm{K}}(0)
\sqrt{\sin \left( \pi \langle n_{z\uparrow}\rangle \right)
\sin \left( \pi \langle n_{z\downarrow}\rangle \right)}
\nonumber\\
&\simeq
T_{\mathrm{K}}(0) \cos \left(\frac{\pi}{2} \langle 
\delta n_f \rangle \right).
\end{align}
As the spin accumulation $\left\langle \delta n_f\right\rangle$
increases, the spin fluctuation is suppressed, which weakens the
Kondo effect. 
$T_{\mathrm{K }}(\delta\epsilon) \simeq 0$ when
$\left\langle \delta n_f\right\rangle \simeq 1$
($\left\langle n_{z\uparrow}\right\rangle \simeq 1$ and
$\left\langle n_{z\downarrow}\right\rangle \simeq 0$).
In this situation, the Kondo effect is very weak because spin
$z\downarrow$ hardly exists in the QD.

The above-mentioned discussion can be generalized to the case of
asymmetric barriers, $V_L\not= V_R$.
The Kondo temperature $T_{\mathrm{K}}(0,\theta,p)$ is given by Eq.\ 
(\ref{Tk02}) in Appendix A. It is obtained
by replacing $p\cos(\theta/2)$
in Eq.\ (\ref{Tk0}) to be
$p\sqrt{\cos^2(\theta/2)+v^2\sin^2(\theta/2)}$.

Now we discuss the conductance $G$ through the QD in the case of 
symmetric barriers. The SBMF calculation yields
\begin{align}
G&=\frac{e^2}{h} \left[\frac{2(1-p^2)}
{1-p^2\cos^2\frac{\theta}{2}}+p^2\sin^2\frac{\theta}{2}
\left(\frac{\delta\epsilon}{T_{\mathrm{K}}(0,\theta,p)}
\right)^2\right]
\nonumber\\
&\quad\times\left(\frac{T_{\mathrm{K}}(\delta\epsilon,\theta,p)}
{T_{\mathrm{K}}(0,\theta,p)}\right)^2.
\label{conduct1}
\end{align}
The conductance is explained as the resonant tunneling
through the Kondo resonant levels at
$\tilde{E}_{z\uparrow/z\downarrow}$ in Eq.\ (\ref{Kondolevel})
with the width of $\tilde{\Gamma}_{\uparrow/\downarrow}$.
With non-magnetic leads ($p=0$), the conductance is simply
given by $G=(2e^2/h)
[1-(\delta\epsilon/2T_{\mathrm{K}}^{(0)})^2]$,
where $T_{\mathrm{K}}^{(0)}$ is the Kondo temperature at
$\delta\epsilon=0$.\cite{com1}
When $\delta \epsilon=0$, the resonant tunneling through the
level of $\tilde{E}_{z\uparrow/z\downarrow}=0$
yields the unitary limit, $G=2e^2/h$. A finite $\delta\epsilon$
splits the resonant level into two, which reduces the conductance.
At $\delta\epsilon \ge 2T_{\mathrm{K}}^{(0)}$, $G=0$ since the
Kondo effect does not take place in the SBMF theory.
With ferromagnetic leads ($p \not= 0$), we expect a
suppression of $G$ below $G=2e^2/h$ owing to
the spin-valve effect,\cite{jul,miy,moo,slo,kon} besides the
splitting of the Kondo resonant levels by finite $\delta \epsilon$.

\begin{figure}[t]
\begin{center}
 \rotatebox{-90}{\includegraphics[scale=0.5]{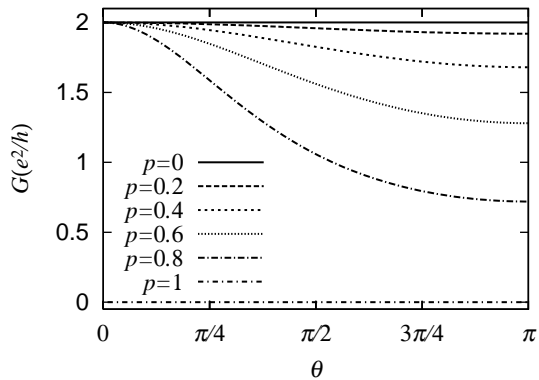}}
\caption{Conductance $G$ through the QD as a function of 
$\theta$, in the absence of spin splitting ($\delta\epsilon=0$).
The spin polarization $p$ in the leads is varied from $0$ to
$1$. The tunnel barriers are symmetric ($v=0$).
}
\label{s-v}
\end{center}
\end{figure}

First, we consider the case of $\delta\epsilon=0$ to
discuss the spin-valve effect only. Then the Kondo resonant
levels are fixed at the Fermi level,
$\tilde{E}_{z\uparrow/z\downarrow}=0$.
Equation (\ref{conduct1}) reduces to
\begin{eqnarray}
G=\frac{2e^2}{h}\frac{1-p^2}{1-p^2\cos^2\frac{\theta}{2}}.
\label{g0}
\end{eqnarray}
$G$ is plotted
as a function of $\theta$ in Fig.\ \ref{s-v}, for various $p$.
When $0<p<1$, the conductance decreases with an increase in $\theta$.
This is due to the spin-valve effect in which the conductance is 
maximal in the P alignment ($\theta=0$) and minimal in the AP
alignment ($\theta=\pi$). Change of $G$, $G(\theta=0)-G(\theta=\pi)$,
is more prominent for larger $p$.
When $p=1$ (half-metallic leads), the conductance always vanishes
for the following reason.
At $\theta = 0$, this is due to the absence of Kondo effect.
At $\theta \not= 0$, this is ascribable to an interference
effect between two spin-components. This is analogous to the 
Fano-type antiresonance in a quantum wire with a side-coupled quantum
dot.\cite{fan,Kobayashi,Sato}
In Appendix B, we examine a non-interacting model in which
a quantum dot is coupled to two ferromagnetic leads with perfect
polarization ($p=1$). We find
a dip of $G$ ($G=0$) when an energy level in the QD
matches the Fermi level in the leads. The dip can be understood
as the Fano-type antiresonance because one spin component in the QD
is coupled to both the leads (corresponding to ``a quantum wire'')
while the other spin component is coupled to one of the leads
(``side-coupled quantum dot'').
In the case of $p=1$ in Fig.\ \ref{s-v}, the dip makes $G=0$ since
the Kondo resonant level always appears at the Fermi level.

\begin{figure}[t]
\begin{center}
 \rotatebox{-90}{\includegraphics[scale=0.34]{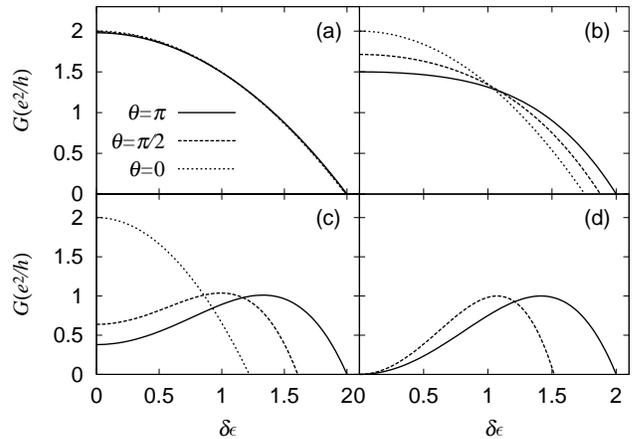}}
\caption{Conductance $G$ through the QD as
a function of spin splitting $\delta\epsilon$.
The spin polarization in the leads
is (a) $p=0.1$, (b) $0.5$, (c) $0.9$ and (d) $1$.
The angle $\theta$ between the magnetic moments in the leads
is $0$, $\pi/2$, and $\pi$.
The tunnel barriers are symmetric ($v=0$).
$\delta\epsilon$ is normalized by
$T_{\mathrm{K}}(\delta\epsilon=0,
\theta=\pi,p)$. Note that
$G=0$ for any $\delta\epsilon$ in the case
of (d) $p=1$ and $\theta=0$. 
}
\label{Gc}
\end{center}
\end{figure}

Second, we examine the conductance $G$ in the presence of
spin splitting $\delta\epsilon$.
In Fig.\ \ref{Gc},
$G$ is plotted as a function of
$\delta\epsilon$, for various $p$ and $\theta$.
$\delta\epsilon$ in the abscissa is normalized by
$T_{\mathrm{K}}(\delta\epsilon=0,\theta=\pi,p)$,
the Kondo temperature at $\delta\epsilon=0$ and $\theta=\pi$ for
each value of spin polarization $p$ in the leads.
When $p$ is small [(a) $p=0.1$, (b) $0.5$], $G$ decreases with
increasing $\delta\epsilon$ and becomes zero at a critical
value of $\delta\epsilon$.
This is due to the splitting
of the Kondo resonant levels for $z\uparrow$ and $z\downarrow$,
as in the case of non-magnetic leads.
When $p$ is large [(c) $p=0.9$, (d) $1$] and $\theta \ne 0$,
the second term in Eq.\ (\ref{conduct1}) makes a non-monotonic
behavior of $G(\delta\epsilon)$.
At $p \sim 1$,
the conductance is suppressed by the dip of Fano-type antiresonance
at $\delta\epsilon=0$. With increasing $\delta\epsilon$, the dip
is shifted from the Fermi level, which increases the conductance.
This results in a maximum of $G$ as a function of $\delta\epsilon$,
in combination with the suppression of $G$ by the splitting of
Kondo resonant levels.

\section{Conclusions}

We have theoretically investigated the Kondo effect in a QD 
coupled to noncollinear ferromagnetic leads,
using the poor man's scaling method and the SBMF theory.

We have used the poor man's scaling procedure in two stages.
In the first stage,
we have reduced the energy scale $D$ until the charge
fluctuation is quenched and renormalized the energy
levels for spin $\uparrow$ and $\downarrow$ in the QD.
We have found a finite spin splitting
$\delta\epsilon=\epsilon_{\downarrow}-\epsilon_{\uparrow}$
in the absence of magnetic field.
This stems from the spin-dependent charge fluctuation.
The spin splitting takes place in an intermediate direction
between the magnetic moments in the two ferromagnets.
$\delta\epsilon \propto
p\sqrt{\cos^2(\theta/2)+v^2\sin^2(\theta/2)}$, where $p$ is the
spin polarization in the leads, $\theta$ is the angle between the
magnetic moments, and $v$ is an asymmetric factor of tunnel
barriers ($-1<v<1$). Hence the spin splitting is always maximal
in the parallel alignment of two ferromagnets ($\theta=0$) and
minimal in the antiparallel alignment ($\theta=\pi$).

We have proceeded to the second stage of the scaling for $D<D_1$,
where a localized spin fluctuates in the QD with fixed spin
splitting $\delta\epsilon$. When $\delta\epsilon$ is irrelevant
($\delta\epsilon \ll T_{\mathrm{K}}$), we have derived 
an analytical expression of the Kondo temperature
$T_{\mathrm{K}}(\theta,p)$. $T_{\mathrm{K}}$ is a decreasing
function with respect to
$p\sqrt{\cos^2(\theta/2)+v^2\sin^2(\theta/2)}$.
When $\delta\epsilon \gg T_{\mathrm{K}}$, the Kondo effect
does not take place.

We have applied the SBMF theory to the renormalized Hamiltonian
by the first stage scaling in which the charge fluctuation in
the energy range of $D_1<D<D_0$ (original band width) has been taken
into account. When $\delta\epsilon=0$, the Kondo temperature
decreases with increasing
$p\sqrt{\cos^2(\theta/2)+v^2\sin^2(\theta/2)}$, in
semi-quantitative agreement with that obtained by the scaling
method. This clearly indicates an applicability of the SBMF theory
to the present problem. 
When $\delta\epsilon$ is comparable with $T_{\mathrm{K}}$, we have
observed a splitting of the Kondo resonance, 
suppression of the Kondo effect, and spin accumulation
in the QD.
$T_{\mathrm{K}}(\delta\epsilon, \theta, p)$ has been evaluated
by the width of the split Kondo resonances.
We have also calculated the conductance $G$ at temperature $T=0$.
$G$ could show a non-monotonic
behavior with increasing $\delta\epsilon$, in contrast to
$T_{\mathrm{K}}$. This is ascribable to an interference effect
between two spin components, analogous to the Fano resonance.

\section*{Acknowledgments}
The authors gratefully acknowledge discussions with Y.\ Utsumi.
This work was partially supported by a Grant-in-Aid for
Scientific Research in Priority Areas ``Semiconductor Nanospintronics''
(No.\ 14076216) of the Ministry of Education, Culture, Sports, Science
and Technology, Japan.

\appendix

\section{Case of asymmetric barriers}

\begin{figure}[t]
\begin{center}
 \includegraphics[scale=0.4]{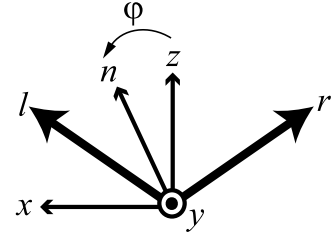}
\caption{Quantization axis $n$ for the QD spin 
in the case of asymmetric barriers, $V_L\not=V_R$.
$l(r)$ indicates the direction of magnetic moment in lead
$L(R)$, which is tilted by $\theta/2$ ($-\theta/2$)
from the $z$ axis in the $z$-$x$ plane (see Fig.\ 1).
The axis $n$ is tilted by $\varphi$ from the $z$ axis 
in the same plane.}
\label{Qaxis}
\end{center}
\end{figure}

In the case of asymmetric barriers, $V_L\not= V_R$,
we can choose the quantization axis for the QD spin,
as in the following.
We define an axis $n$ which is tilted by $\varphi$ 
from the $z$ axis in the $z$-$x$ plane as shown 
in Fig.\ \ref{Qaxis}. We denote $n\uparrow/n\downarrow$
for spin-up/down in the direction.
Fermion operators in the QD,  $d_{l\sigma}$, $d_{r\sigma}$,
are related to $d_{n\sigma}$ by 
\begin{align}
&\left\{
\begin{array}{l}
d_{l\uparrow}=\cos\frac{\theta-2\varphi}{4}d_{n\uparrow}
+\sin\frac{\theta-2\varphi}{4}
d_{n\downarrow},\\ 
d_{l\downarrow}=-\sin\frac{\theta-2\varphi}{4}d_{n\uparrow}
+\cos\frac{\theta-2\varphi}{4}
d_{n\downarrow},
\end{array}
\right.
\\
&
\left\{
\begin{array}{l}
d_{r\uparrow}=\cos\frac{\theta+2\varphi}{4}d_{n\uparrow}
-\sin\frac{\theta+2\varphi}{4}
d_{n\downarrow},\\
d_{r\downarrow}=\sin\frac{\theta+2\varphi}{4}d_{n\uparrow}
+\cos\frac{\theta+2\varphi}{4}
d_{n\downarrow}.
\end{array}
\right.
\end{align}
We consider a following transformation for conduction electrons
with $\epsilon \le \epsilon_{k\uparrow}, \epsilon_{k'\downarrow}
< \epsilon+d\epsilon$.
For $\rho_{\downarrow} d\epsilon$ of majority spins,
\begin{eqnarray}
\left\{
\begin{array}{l}
a_{k\uparrow}=A_1c_{k\uparrow}-A_2\tilde{c}_{k\uparrow}
+A_3c_{k'\downarrow}-A_4\tilde{c}_{k'\downarrow},\\
a_{k\downarrow}=-A_3c_{k\uparrow}-A_4\tilde{c}_{k\uparrow}
+A_1c_{k'\downarrow}+A_2\tilde{c}_{k'\downarrow},\\
\bar{a}_{k\uparrow}=A_2c_{k\uparrow}+A_1\tilde{c}_{k\uparrow}
+A_4c_{k'\downarrow}+A_3\tilde{c}_{k'\downarrow},\\
\bar{a}_{k\downarrow}=A_4c_{k\uparrow}-A_3\tilde{c}_{k\uparrow}
-A_2c_{k'\downarrow}+A_1\tilde{c}_{k'\downarrow}.
\end{array}
\right.
\label{UT1}
\end{eqnarray}
For the rest of the majority spins,
\begin{eqnarray}
\left\{
\begin{array}{l}
a_{k\uparrow}=\frac{1}{\sqrt{A_1^2+A_2^2}}
(A_1c_{k\uparrow}-A_2\tilde{c}_{k\uparrow}),\\
a_{k\downarrow}=\frac{1}{\sqrt{A_3^2+A_4^2}}
(-A_3c_{k\uparrow}-A_4\tilde{c}_{k\uparrow}),
\end{array}
\right.\label{UT2}
\end{eqnarray}
The coefficients are given by
\begin{align}
\left\{
\begin{array}{l}
A_1=\cos\frac{\theta}{4}\cos\frac{\varphi}{2}
+v\sin\frac{\theta}{4}\sin\frac{\varphi}{2},\\
A_2=v_{LR}\sin\frac{\theta}{4}\sin\frac{\varphi}{2},\\
A_3=\cos\frac{\theta}{4}\sin\frac{\varphi}{2}
-v\sin\frac{\theta}{4}\cos\frac{\varphi}{2},\\
A_4=v_{LR}\sin\frac{\theta}{4}\cos\frac{\varphi}{2}.
\end{array}
\right.
\end{align}
$c_{k\sigma}$ and $\tilde{c}_{k}$ are defined as
\begin{align}
\left\{
\begin{array}{l}
c_{k\uparrow}=\frac{1}{\mathcal{V}}(V_Lc_{Lk,l\uparrow}
+V_Rc_{Rk,r\uparrow}),\\
\tilde{c}_{k\uparrow}=\frac{1}{\mathcal{V}}(-V_Rc_{Lk,l\uparrow}
+V_Lc_{Rk,r\uparrow}),\\
c_{k\downarrow}=\frac{1}{\mathcal{V}}(V_Lc_{Lk,l\downarrow}+
V_Rc_{Rk,r\downarrow}),\\
\tilde{c}_{k\downarrow}=\frac{1}{\mathcal{V}}(V_Rc_{Lk,l\downarrow}
-V_Lc_{Rk,r\downarrow}),
\end{array}
\right.
\end{align}
where $\mathcal{V} = \sqrt{V_L^2+V_R^2}$, 
$v_{LR} = 2V_LV_R/\mathcal{V}^2$, and
$v$ is the asymmetric factor defined by Eq.\ (\ref{asymmetric-f}).
Then the tunnel Hamiltonian $H_{\mathrm{T }}$ 
[the last term in Eq.\ (\ref{anderson})] 
is rewritten as 
\begin{eqnarray}
H_{\mathrm{T }}=\mathcal{V}\sum_{\sigma}
(A_{\sigma}^{\dagger}d_{n\sigma}
+\textrm{h.c.}),
\end{eqnarray}
where
\begin{eqnarray}
\left\{
\begin{array}{l}
A_{\uparrow}=\sum_{k}{}'a_{k\uparrow}
+\sqrt{A_1^2+A_2^2}\sum_{k}{}''a_{k\uparrow},
\\
A_{\downarrow}=\sum_{k}{}'a_{k\downarrow}
+\sqrt{A_3^2+A_4^2}\sum_{k}{}''a_{k\downarrow}.
\end{array}
\right.
\end{eqnarray}
The modes of $\bar{a}_{k\uparrow}$ and $\bar{a}_{k\downarrow}$ are
decoupled from the QD.

Although Eq.\ (\ref{UT1}) is the unitary transformation with
arbitrary coefficients, Eq.\ (\ref{UT2}) is not.
From the condition $[A_{\uparrow},A_{\downarrow}^{\dagger}]_+=0$, 
that is, $A_1A_3=A_2A_4$,  
we determine $\varphi$ (direction of axis $n$) so that
Eq.\ (\ref{UT2}) becomes the unitary transformation.
It is given by
\begin{eqnarray}
\tan\varphi =v\tan\frac{\theta}{2}.
\label{phi}
\end{eqnarray}
The quantization axis $n$ leans toward the direction of
magnetic moment with larger tunnel coupling
from the $z$ direction.

We calculate the spin splitting,
$\delta\epsilon=\epsilon_{n\downarrow}-\epsilon_{n\uparrow}$,
in the same way as in Sec.\ III.\ A.
In the most right-hand side of Eq.\ (\ref{split1}),
$p\cos(\theta/2)$ should be replaced by
$p\cos(\theta/2)/\cos\varphi=
p\sqrt{\cos^2(\theta/2)+v^2\sin^2(\theta/2)}$.
This yields
\begin{eqnarray}
\delta\epsilon=\frac{\Delta}{\pi}
p\sqrt{\cos^2\frac{\theta}{2}+v^2\sin^2\frac{\theta}{2}}
\ln\frac{D_0}{D_1}
-E_{\mathrm{Z }}.
\label{g-delep}
\end{eqnarray}
In the case of asymmetric tunnel barriers ($v \ne 0$),
we observe a finite spin splitting even in the AP alignment,
$\delta\epsilon=(\Delta/\pi)p|v|\ln(D_0/D_1)$,
when $E_{\mathrm{Z}}=0$.

The Kondo temperatures are obtained using the scaling method
(Sec.\ III.\ B) or by the SBMF theory (Sec.\ IV).
Replacing $p\cos(\theta/2)$ in Eqs.\ (\ref{KondoTemp}) and
(\ref{Tk0}) by
$p\sqrt{\cos^2(\theta/2)+v^2\sin^2(\theta/2)}$,
they are generalized to
\begin{align}
&T_{\mathrm{K}}(\theta,p)
=D_1\exp\left[-\frac{1}{(\rho_{\uparrow}+
\rho_{\downarrow})J}\right.\nonumber\\
&\quad\quad\quad\quad\left.\times \frac{\textrm{arctanh}
\left(p\sqrt{\cos^2\frac{\theta}{2}+v^2\sin^2\frac{\theta}{2}}\right)}
{p\sqrt{\cos^2\frac{\theta}{2}+v^2\sin^2\frac{\theta}{2}}}\right],
\label{KondoTemp2}
\end{align}
and 
\begin{align}
&T_{\mathrm{K }}(0,\theta,p)=
D_1\exp\left[-\frac{\pi|\epsilon|}{\Delta}\right.\nonumber\\
&\left.-p\sqrt{\cos^2\frac{\theta}{2}+v^2\sin^2\frac{\theta}{2}}
\textrm{arctanh}
\left(p\sqrt{\cos^2\frac{\theta}{2}
+v^2\sin^2\frac{\theta}{2}}\right)\right],
\label{Tk02}
\end{align}
respectively.
With an increase in the asymmetric factor $|v|$, the spin fluctuation
in the QD is suppressed and accordingly the Kondo effect is weakened.

\section{Non-interacting model with $p=1$}

In Appendix B, we examine a non-interacting model in
which a QD is coupled to noncollinear ferromagnetic leads.
We assume that the leads are half-metals
($\rho_{\downarrow}=0$) and thus the spin polarization is $p=1$.
With $U=0$, the Hamiltonian (\ref{anderson}) reads
\begin{eqnarray}
H&=&\sum_{k}\epsilon_{k\uparrow}c_{Lk,l\uparrow}^{\dagger}c_{Lk,l\uparrow}+
\sum_{k}\epsilon_{k\uparrow}c_{Rk,r\uparrow}^{\dagger}c_{Rk,r\uparrow}
\nonumber\\
&&+\sum_{\sigma}\epsilon_{0}d_{z\sigma}^{\dagger}d_{z\sigma}
+H_{\mathrm{T}}.
\end{eqnarray}
Here, the tunnel Hamiltonian is written as
\begin{eqnarray}
H_{\mathrm{T}}=
\sum_{k\sigma}(V_{L}c_{Lk,l\sigma}^{\dagger}d_{l\sigma}
+V_{R}c_{Rk,r\sigma}^{\dagger}d_{r\sigma}+\mathrm{h.c.}),
\label{tunnelB}
\end{eqnarray}
or 
\begin{eqnarray}
H_{\mathrm{T}}=
 \mathcal{V} \sum_{\sigma}
(A_{\sigma}^{\dagger}d_{n\sigma}+\mathrm{h.c.}),
\end{eqnarray}
using the quantum axis $n$ in Eq.\ (\ref{phi}) in Appendix A.

\begin{figure}[t]
\begin{center}
 \rotatebox{-90}{\includegraphics[scale=0.5]{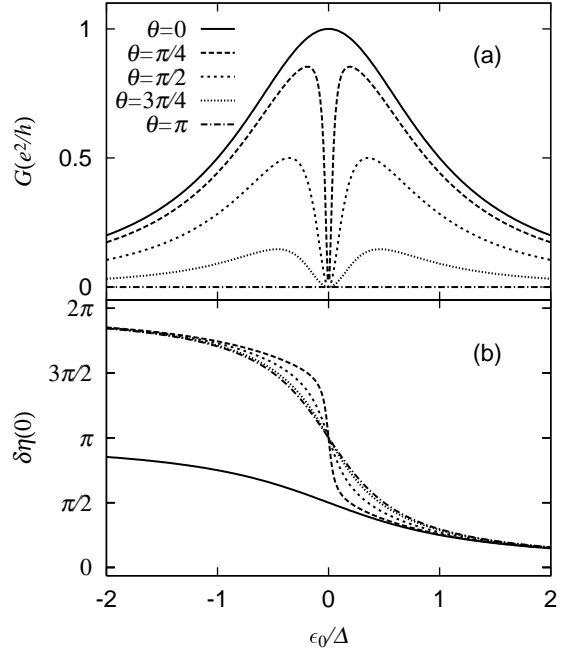}}
\caption{Calculated results using a non-interacting model in
which a QD is coupled to two noncollinear ferromagnetic leads
with perfect polarization ($p=1$).
(a) Conductance $G$ and (b) phase shift at the Fermi level
$\delta\eta(0)$, as functions of QD level $\epsilon_0$.
$\epsilon_0$ is normalized by the tunnel coupling strength
$\Delta$ given by Eq.\ (\ref{tunnel-strength}).
The tunnel barriers are symmetric ($v=0$).
The angle $\theta$ between the magnetic moments in the leads
is shown in panel (a).}
\label{g3}
\end{center}
\end{figure}

By virtue of the quantum axis $n$, the Green functions in the QD
can be written in a diagonal form,
\begin{align}
&\hat{G}_{n\sigma n\sigma}(\omega)
=\frac{1}{\omega-\epsilon_0
+\mathrm{i}\Gamma_{n\sigma}},\\
&\hat{G}_{n\uparrow n\downarrow}(\omega)
=\hat{G}_{n\downarrow n\uparrow}(\omega)=0,
\end{align}
where
\begin{eqnarray}
\Gamma_{n\uparrow/n\downarrow} &=&
\frac{\Delta}{2}
\left[ 1 \pm \frac{\cos(\theta/2)}{\cos\varphi}
\right]
\nonumber \\
&=&
\frac{\Delta}{2}
\left[ 1 \pm \sqrt{\cos^2(\theta/2)+v^2\sin^2(\theta/2)}
\right]
\label{GammaB}
\end{eqnarray}
are the level broadenings for spin $n\uparrow / n\downarrow$
in the QD.
The T-matrix $\hat{T}$ through the QD is given by
\begin{align}
\left\langle Rk,r\uparrow\right|\hat{T}
\left|Lk,l\uparrow\right\rangle
&=V_LV_R
\biggl[
\cos\frac{\theta-2\varphi}{4}\cos\frac{\theta+2\varphi}{4}
\hat{G}_{n\uparrow n\uparrow}(\omega)\nonumber\\
&\quad-\sin\frac{\theta-2\varphi}{4}\sin\frac{\theta+2\varphi}{4}
\hat{G}_{n\downarrow n\downarrow}(\omega) \biggr]
\nonumber\\
&=\frac{V_LV_R(\omega-\epsilon_0)\cos\frac{\theta}{2}}
{(\omega-\epsilon_0
+\mathrm{i}\Gamma_{n\uparrow})(\omega-\epsilon_0
+\mathrm{i}\Gamma_{n\downarrow})},
\label{eq:T-mtx}
\end{align}
with $\omega$ being the energy of incident electrons.
Using the T-matrix, we obtain the
conductance $G$ through the QD
\begin{align}
G&=\frac{e^2}{h}(2\pi\rho_{\uparrow})^2
|\left\langle Rk,r\uparrow\right|\hat{T}
\left|Lk',l\uparrow\right\rangle|^2 \biggr|_{\omega=0}
\nonumber\\
&=\frac{e^2}{h}\frac{4\Gamma_{L}\Gamma_{R}
\epsilon_0^2\cos^2\frac{\theta}{2}}
{(\epsilon_0^2+\Gamma_{n\uparrow}^2)
(\epsilon_0^2+\Gamma_{n\downarrow}^2)}.
\label{conduct3}
\end{align}
$\Gamma_{L}=\pi \rho_{\uparrow}V_{L}^2$ and
$\Gamma_{R}=\pi \rho_{\uparrow}V_{R}^2$ are the strength
of the tunnel coupling to leads $L$ and $R$, respectively.
$\Delta=\Gamma_{L}+\Gamma_{R}$ in Eq.\ (\ref{tunnel-strength})
for $\rho_{\downarrow}=0$.

In Fig.\ \ref{g3}(a), we present the conductance $G$ as a
function of QD level $\epsilon_0$.
In the P alignment ($\theta=0$),
$\Gamma_{n\uparrow}=\Delta$ and $\Gamma_{n\downarrow}=0$ in Eq.\
(\ref{GammaB}). Then Eq.\ (\ref{conduct3}) is reduced to
$G=(e^2/h) \cdot 4\Gamma_{L}\Gamma_{R}/
(\epsilon_0^2+\Delta^2)$, which indicates the usual resonant
tunneling through QD level $\epsilon_0$
by one spin-component [solid line in Fig.\ \ref{g3}(a)].
When $0<\theta<\pi$, however, Eq.\ (\ref{conduct3}) yields
$G=0$ at $\epsilon_0=0$. We observe a dip of $G$ at
$\epsilon_0=0$ in Fig.\ \ref{g3}(a). 
In the AP alignment ($\theta=\pi$),
$G=0$ for any value of $\epsilon_0$ since we are considering
half-metallic leads with $\rho_{\downarrow}=0$.

The dip of the conductance at $0<\theta<\pi$ is attributable
to an interference effect between two spin components. 
Consider the tunnel coupling between lead $L$ and QD in
tunnel Hamiltonian (\ref{tunnelB}). The QD state
$| l \uparrow \rangle$ is coupled to the lead $L$ by $V_L$,
whereas $| l \downarrow \rangle$ is decoupled since
$\rho_{\downarrow}=0$ in the lead. On the other hand,
both the QD states $| l \uparrow \rangle$ and
$| l \downarrow \rangle$ are coupled to the
lead $R$ by $V_R \cos (\theta/2)$ and $-V_R \sin (\theta/2)$,
respectively, because the QD state $| r \uparrow \rangle$
is coupled to the lead $R$ by $V_R$ and the QD state is
rewritten as
$| r \uparrow \rangle= \cos (\theta/2) | l \uparrow \rangle
-\sin (\theta/2) | l \downarrow \rangle$. 
This situation is analogous to that of a quantum wire
(lead $L$ -- QD state $| l \uparrow \rangle$ --lead $R$)
with a side-coupled QD
(QD state $| l \downarrow \rangle$ --lead $R$), in which
the Fano-type antiresonance has been observed.\cite{Kobayashi,Sato}
In general, an interference between a localized state and
a continuum of states results in an asymmetric Fano resonance,
$G \propto (\tilde{\epsilon}_0+q)^2/(\tilde{\epsilon}_0^2+1)$, where
$\tilde{\epsilon}_0$ is the resonant level normalized by the
resonant width.\cite{fan} In a geometry of a quantum wire with
a side-coupled QD, the Fano factor is $q=0$ which makes a dip
at $\tilde{\epsilon}_0=0$.

An alternative explanation for $G=0$ at $\epsilon_0=0$
is given in terms of phase shift.
Using the T-matrix in Eq.\ (\ref{eq:T-mtx}),
the phase shift $\delta\eta(\omega)$ is defined by
\begin{align}
\delta\eta(\omega)&=\arg\left\langle Rk,r\uparrow\right|\hat{T}
\left|Lk,l\uparrow\right\rangle \nonumber\\
&=\arctan\frac
{(\epsilon_0-\omega)(\Gamma_{n\uparrow}+\Gamma_{n\downarrow})}
{(\epsilon_0-\omega)^2-\Gamma_{n\uparrow}\Gamma_{n\downarrow}}.
\end{align}
Using Eq.\ (\ref{GammaB}),
the conductance $G$ in Eq.\ (\ref{conduct3}) is rewritten as
\begin{eqnarray}
G=\frac{e^2}{h}\frac{4\Gamma_L\Gamma_R}{(\Gamma_L+\Gamma_R)^2}
   \cos^2 \frac{\theta}{2} \sin^2 \delta\eta(0),
\label{conduct4}
\end{eqnarray}
where $\delta\eta(0)$ is the phase shift for electrons at the
Fermi level, $\omega=0$.

Next, we express the number of electrons in the QD
$\langle n_d\rangle$ in terms of
the phase shift $\delta\eta(0)$ (Friedel sum rule).
Using the local density of states in the QD
\begin{eqnarray}
\rho_{d}(\omega) \equiv
-\sum_{\sigma} \frac{1}{\pi} \mathrm{Im}\hat{G}_{n\sigma n\sigma}
(\omega)
=-\frac{1}{\pi}\frac{d\delta\eta(\omega)}{d\omega},
\end{eqnarray}
we obtain
\begin{eqnarray}
\langle n_d\rangle = \int_{-\infty}^{0}\rho_{d}(\omega )d\omega
= \frac{\delta\eta(0)}{\pi}.
\label{Friedel}
\end{eqnarray}
This sum rule is different by a factor of one half from
the conventional Friedel sum rule in the case of non-magnetic leads.

Figure \ref{g3}(b) shows the phase shift at the Fermi level
$\delta\eta(0)$, as a function of QD level $\epsilon_0$.
At $\epsilon_0=0$, $\delta\eta(0)=\pi/2$
when $\theta=0$ and $\delta\eta(0)=\pi$ when $\theta \not=0$,
corresponding to $G=(2e^2/h)\cdot
4\Gamma_L\Gamma_R/(\Gamma_L+\Gamma_R)^2$ and $G=0$
in Fig.\ \ref{g3}(a), respectively.
This can be explained using the sum rule of
Eq.\ (\ref{Friedel}). In the half-filling case of $\epsilon_0=0$,
the number of electrons should be unity when $\theta \not=0$.
Then Eq.\ (\ref{Friedel}) yields $\delta\eta(0)=\pi$
and hence $G=0$ by Eq.\ (\ref{conduct4}).
In a special case of $\theta=0$, the QD state
$|z\downarrow \rangle$ is always empty since it is completely
decoupled from two half-metallic leads. Then $\langle n_d\rangle=1/2$
at $\epsilon_0=0$ and thus $\delta\eta(0)=\pi/2$,
which leads to $G=(e^2/h)\cdot
4\Gamma_L\Gamma_R/(\Gamma_L+\Gamma_R)^2$.


\begin{thebibliography}{0}
\bibitem{hew} A.\ C.\ Hewson, The Kondo Problem to Heavy Fermions
(Cambridge University Press, Cambridge, UK, 1993).
\bibitem{exp1} D.\ Goldhaber-Gordon, H.\ Shtrikman, D.\ Mahalu, D.\ 
Abusch-Magder, U.\ Meirav, and M.\ A.\ Kastner Nature
{\bf 391}, 156 (1998). 
\bibitem{exp2} S.\ Cronenwett, T.\ H.\ Oosterkamp, and
L.\ P.\ Kouwenhoven, Science {\bf 281}, 540 (1998).
\bibitem{spintronics}
I.\ \v{Z}uti\'c, J.\ Fabian, and S.\ Das Sarma,
Rev.\ Mod.\ Phys.\ {\bf 76}, 323 (2004).
\bibitem{jul} M.\ Julliere, Phys.\ Lett.\ {\bf 54A}, 225 (1975).
\bibitem{miy} T.\ Miyazaki and N.\ Tezuka, J.\ Magn.\ Magn.\ Mater.\ 
{\bf 139}, L231 (1995).
\bibitem{moo} J.\ S.\ Moodera and L.\ R.\ Kinder, J.\ Appl.\ Phys.\ 
{\bf 79}, 4724 (1996).
\bibitem{slo} J.\ C.\ Slonczewski, Phys.\ Rev.\ B {\bf 39}, 6995 (1989).
\bibitem{kon} J.\ K\"{o}nig, and J.\ Martinek, Phys.\ Rev.\ Lett.\ {\bf 90}, 
166602 (2003).
\bibitem{bra} M.\ Braun, J.\ K\"{o}nig, and J.\ Martinek, Phys.\ Rev.\ 
B.\ {\bf 70}, 
195345 (2004).
\bibitem{ser} N.\ Sergueev, Q.-F.\ Sun, H.\ Guo, B.\ G.\ Wang, and J.\
Wang, Phys.\ Rev.\ B {\bf 65}, 165303 (2002).
\bibitem{bul} B.\ R.\ Bulka and S.\ Lipi\'{n}ski,
 Phys.\ Rev.\ B 67, 024404 (2003).
\bibitem{mar1} J.\ Martinek, Y.\ Utsumi, H.\ Imamura, J.\ Barna\'{s}, S.\
Maekawa, J.\ K\"{o}nig, and G.\ Sch\"{o}n, Phys.\ Rev.\ Lett.\ {\bf 91},
127203 (2003).

\bibitem{jma} J.\ Ma, B.\ Dong, and X.\ L.\ Lei, cond-mat/0212645.
\bibitem{don} B.\ Dong, H.\ L.\ Cui, S.\ Y.\ Liu, and X.\ L.\ Lei,
 J.\ Phys.:Condens.\ Matter {\bf 15}, 8435 (2003).
\bibitem{mar2} J.\ Martinek, M.\ Sindel,
 L.\ Borda, J.\ Barna\'{s}, J.\ K\"{o}nig,
G.\ Sch\"{o}n, and J.\ von Delft, Phys.\ Rev.\ Lett.\ {\bf 91}, 247202
(2003).
\bibitem{cho} M.-S.\ Choi, D.\ S\'{a}nchez, and R.\ L\'{o}pez,
 Phys.\ Rev.\ Lett.\ {\bf 92}, 056601 (2004).
\bibitem{uts} Y.\ Utsumi, J.\ Martinek, G.\ Sch\"{o}n, H.\ Imamura, S.\
Maekawa, Phys.\ Rev.\ B {\bf 71}, 245116 (2005).
\bibitem{zha} P.\ Zhang, Q.-K.\ Xue, Y.-P.\ Wang, and X.\ C.\ Xie, Phys.\
Rev.\ Lett.\ 89, 286803 (2002).
\bibitem{lop} R.\ L\'{o}pez and D.\ S\'{a}nchez,
 Phys.\ Rev.\ Lett.\ 90, 116602 (2003).
 
\bibitem{swi} R.\ \'{S}wirkowicz, M.\ Wilczy\'{n}ski, M.\ Wawrzyniak, 
and J.\ Barna\'{s}, Phys.\ Rev.\ B {\bf 73}, 193312 (2006). 
\bibitem{pas} A.\ N.\ Pasupathy, R.\ C.\ Bialczak, J.\ Martinek, J.\ E.\
Grose, L.\ A.\ K.\ Donev, P.\ L.\ McEuen, and D.\ C.\ Ralph,
Science {\bf 306}, 85 (2004).
\bibitem{nyg} J.\ Nyg\r{a}rd, W.\ F.\ Koehl, N.\ Mason, L.\ DiCarlo, and C.\
M.\ Marcus, cond-mat/0410467.
\bibitem{and} P.\ W.\ Anderson, J.\ Phys.\ C {\bf 3}, 2439 (1970).

\bibitem{hal} F.\ D.\ M.\ Haldane, Phys.\ Rev.\ Lett.\ {\bf 40}, 416 (1978).

\bibitem{col} P.\ Coleman, Phys.\ Rev.\ B {\bf 29}, 3035 (1984). 

\bibitem{com5}
This spin splitting has been discussed in a different
context.\cite{kon,bra} The Zeeman-like splitting by the effective
exchange field in Eq.\ (3.9) in Ref.\ \onlinecite{bra} coincides
with our formula, Eq.\ (\ref{g-delep0}), if $D_1$ is replaced by
$|\epsilon_0|$.

\bibitem{com3}
   When $\delta\epsilon=0$, Eq.\ (\ref{eqd1}) is reduced to
   $-D_1 \approx \epsilon_{0}+[\Delta/(2\pi)] \ln (D_0/D_1)$.
   Hence $D_1$ is independent of $p\cos(\theta/2)$.
   We estimate that $D_1/\Delta \approx 1.43$ in 
   the experimental situation of $D_0/\Delta=50$ and
   $\epsilon_0/\Delta=-2$.\cite{pas}


\bibitem{com4}
In the SBMF theory, the Kondo resonant levels appear at
$\tilde{E}_{\sigma}=\epsilon_{\sigma}+\lambda$
for spin $\sigma=z\uparrow/z\downarrow$. They are
shifted from $\epsilon_{\sigma}$ by the same amount
$\lambda$. In other words, the spin splitting is not
renormalized: $\tilde{E}_{z\downarrow}-\tilde{E}_{z\uparrow}
=\epsilon_{z\downarrow}-\epsilon_{z\uparrow}$.
This is because the charge fluctuation is neglected in this theory.

\bibitem{com6}
The sudden disappearance of the Kondo effect
should be attributed to the neglect of
a small charge fluctuation in the energy range of $D<D_1$
in the SBMF theory. It would be corrected by an advanced method,
e.g.\ non-crossing approximation. The SBMF theory is reliable
when the spin fluctuation is dominant and hence the Kondo
temperature is not too small.

\bibitem{com1}
M.\ Eto and Yu.\ V.\ Nazarov, Phys.\ Rev.\ B {\bf 64},
085322 (2001).

\bibitem{fan} U.\ Fano, Phys.\ Rev.\ {\bf 124}, 1866 (1961).
\bibitem{Kobayashi}
K.\ Kobayashi, H.\ Aikawa, A.\ Sano, S.\ Katsumoto, and Y.\ Iye,
Phys.\ Rev.\ B {\bf 70}, 035319 (2004).
\bibitem{Sato}
M.\ Sato, H.\ Aikawa, K.\ Kobayashi, S.\ Katsumoto, and Y.\ Iye,
Phys.\ Rev.\ Lett.\ {\bf 95}, 066801 (2005).


\end{thebibliography}
\end{document}